\documentstyle[epsf,prl,twocolumn,aps]{revtex}
\begin{document}
\title{   Exploring Level Statistics from  
 Quantum Chaos to Localization with  
 the Autocorrelation Function
of Spectral Determinants}
\author{Stefan Kettemann}
\address{ Max- Planck Institut f\" ur Physik Komplexer Systeme\\
 N\" othnitzerstr. 38, 01187 Dresden, Germany}
\date{\today}
\maketitle

\begin{abstract}
The  autocorrelation function of  spectral determinants (ASD)
 is used to characterize the discrete spectrum 
of a phase coherent 
 quasi- 1- dim., disordered  metal wire as a function of its length L
at finite, weak magnetic field. An  analytical function
 is obtained, depending only  on the  
 dimensionless  conductance $ g =  \xi/L$,
 where $\xi$ is the localization length, 
the scaled frequency $x = \omega/\Delta$, where $\Delta$ is the average
 level spacing of the wire, and the global symmetry of the system.  
  A  metal-insulator crossover  
is observed, showing 
 that information on
 localization is contained in  the disorder averaged ASD.

Pacs- numbers:  72.15.Rn,73.20.Fz,73.23.-b
\end{abstract}

\section{Introduction}

 Since the pioneering work of Anderson on localization\cite{anderson58},
 it was realized by Mott and Twose\cite{mott},
 that all states in an infinite 1- dimensional chain are localized
 at arbitrary disorder strength, at zero temperature.
  They could show  that solutions of the
 corresponding  Schroedinger equation  
 at points  at a distance $x$ are  with 
 a probability $ P = \exp ( - x/(4 l))$ in resonance,
 where $l $ is the mean free path.
 Thus, its exponential decrease gives a localization length of the order
 of $l$. 
 Later on Thouless  argued that this statement can be extended
 to a thin disordered wire, and he found that 
 the localization length is given by $\xi =(\pi/3)  M l$
 where $l$ is the elastic mean free path, and $ M = S k_F^2/\pi $
 the number of transverse channels,
 with the Fermi wave vector $k_F$, and the cross section of the wire, 
$S$\cite{thouless}. 
 This was proven rigourously for a matrix of $M$ chains, when $M$ is small,
 by Anderson et al., Weller et al., and then by Dorokhov 
 by solution of a Fokker- Planck Equation\cite{fok}, calculating the 
 transmission probability through the wire.
 In the limit of a thin wire, where the motion of the electrons is
 diffusive in all directions on small length scales,
 the proof of complete localization at zero temperature was
 given by Efetov and Larkin with a field theoretical method,
 obtaining an exponential decrease of the density-density correlation 
 function in space, 
 in the zero frequency limit.  
    They discovered in addition that the localization length 
  depends on the global symmetry of the wire\cite{larkin,ef}.
 The localization length was found to be given by
 $L_c = (1/3) \beta M l$, where $\beta =1, 2, 4$
 for orthogonal, unitary, and symplectic symmetry, 
 which corresponds to no magnetic field, weak magnetic field, 
 and strong spin- orbit interaction, respectively.
 
 The density autocorrelation function was recently studied 
 for the total spectrum as a function of the length of the 
 mesoscopic wire by Altland and Fuchs \cite{alt}.
 Because of the complexity 
  of the problem, they did not obtain a closed analytical expression
 for arbitrary frequency, but succeeded to do a numerical analysis
 in the unitary regime.
 New information on the level statistics of the wire as a function 
 of its length was found in the metal-insulator-crossover regime.

 In this article we argue that in order to study the level statistics,
 it is enough to calculate the simpler ASD.  This function contains
  information on the spectrum, 
 but its complexity is reduced so that
 it can be calculated analytically more easily. 
   We will show that it provides a usefull tool to study localization,
 and could be used  
in situations which have been unaccessible to other analytical methods.

 The article is organized as follows. In the first part 
  the characterization of level statistics by an autocorrelation function 
 is reviewed, and  the ASD is defined.
 In the second part 
the 
 result for the disorder averaged ASD 
 for a quasi-1-dimensional wire in a weak magnetic field 
is presented and discussed 
 for various regimes in the frequency-length plane.
 Information on the metal-insulator-crossover is obtained.
  We conclude with a discussion of the results and the potential
 of the ASD as a new tool to study Anderson localization. 

\section{ Level Statistics as Characterized by an Autocorrelation Function}

 The crossover from a 
metal to an 
insulator in a finite coherent, disordered metal particle is accompanied 
by a change in the
  statistics of the discrete energy levels\cite{ef}. 
 This can be studied by calculating a disorder averaged 
autocorrelation function   
between two energies at a distance
 $\omega$ in the energy level spectrum. Then, 
considering a quasi-1-dim. disordered metal wire
with crossection $S$,
a map 
 as a function of its length $L$ and the energy $\omega$
 can be drawn as in Fig. 1.
 Here, $\Delta = 1/(\nu S L)$ is the total mean level spacing
 with the average density of states $\nu = m k_F/(\pi^2 \hbar^2)
 = 3 n /(2 \epsilon_F)$. $n = N/V $ is the number $N$ of
 electrons 
 per volume $V = S L$. $ \epsilon_F$ is the Fermi 
 energy,  and $m$ the electron mass.
 $\Delta_c = 1/(\nu S \xi)$ is the local mean level spacing, when
 the length of the wire, $L$, exceeds its localization length
$\xi$.
\begin{figure}[bhp]\label{fig1}
\centerline{\hbox{\epsfbox{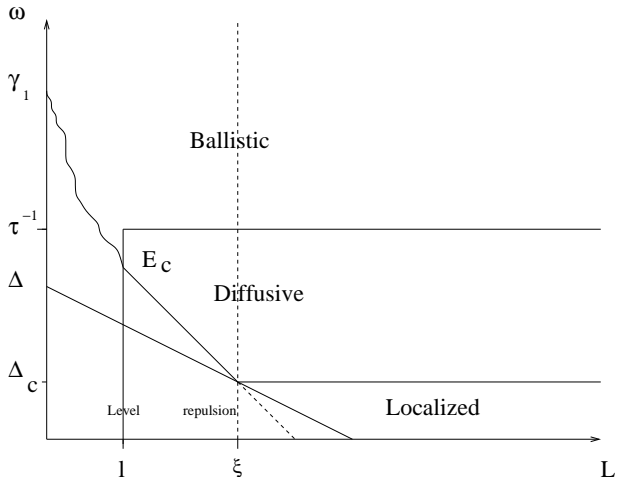}}}
\caption{The map of the energy level spectrum of a quasi-1-dim.
conductor as a function of its length L and characterized
by an autocorrelation function at two points of the spectrum, a distance
 $\omega$ away from each other.}
\end{figure}  
 The Thouless energy $ E_c = \pi^2 \hbar  D/L^2$
 is defined through  the diffusion time across the length L,
 $t_c = 2 \pi  \hbar/E_c$
 when the diffusion is free, as obtained from the classical diffusion
 equation $ \partial_t n = D \partial_x^2 n$, where $n$ is 
 the electron density.
 The classical Diffusion constant D in 3 dimensions is 
 related to the elastic mean free time $\tau$ by
 $ D = v_F^2 \tau/3$.
 $\gamma_1$ is the energy which 
 limits the universal (ergodic)
 regime of nonintegrable ballistic 
 quantum billiards \cite{agam}, \cite{gefen}. Since $\gamma_1$
 depends on the exact boundary conditions, it may change as a function of L
 in a continous, but nonmonotonous way as indicated in Fig.~1. 
 This map has been  explored  by considering  the
 autocorrelation function of density of states,  see \cite{ef,alt}
 and references therein.

 Here, we will restrict us to  
the ASD, 
 as defined by
\begin{equation}
C(\omega) = \frac{\bar{C}(\omega)}{\bar{C}(0)},
\end{equation}
where
\begin{equation}
\bar{C}(\omega) = < det( E + \frac{1}{2}\omega - H) det ( E- \frac{1}{2}\omega -H ) >.
\end{equation}
   $H$ is the hamiltonian of the considered system, 
 and $E$ a central energy.
 It contains only information about the 
 spectrum, not on the wave functions of the mesoscopic system.

 This function was recently obtained for the Wigner- Dyson
 random matrices (GUE\cite{and}, GOE\cite{us}, CUE, COE\cite{haake} and crossover\cite{us}) 
 and successfully used to characterize  the spectrum of nonintegrable
  quantum systems, like a Sinai billiard\cite{us}, for energies 
$ \omega < \gamma_1$,  where the average over the energy $E$ 
 had to be done in order to obtain a universal function.

 In the following , it is obtained  for 
 free electrons
  in a random potential in a finite system, which can be used
 for the  study of a disordered mesoscopic 
 metal  of which at least one dimension exceeds
 the mean free path l.
 The ASD can be calculated 
  analytically and shown to agree with the one for 
 Wigner-Dyson random matrices,
 when the dimensions of the system do not exceed a
 localization length $\xi$. This is expected, since a disordered  
metal particle  
is an example  
of a nonintegrable physical system, and should therefore have the 
same energy level statistics as f.e. the Sinai billiard,
 for frequencies not exceeding the Thouless energy, 
 the ergodic regime, where a particle has time enough to
cover the whole sample uniformly\cite{bohigas}.

Here, we will derive the ASD   for the more general
 case, when one dimension may exceed the localization length $\xi$,
 a quasi-1-dim. conductor, of length $L$ and crossection
 $S$ where the number of transverse channels is much larger
 than 1, $M= S k_F^2/\pi \gg 1$, in order to find out if this
 relatively simple function does contain information about
 the metal-insulator crossover.

\section{From Quantum Chaos to Localization:  the frequency- length plane}  
 
The hamiltonian is given by,
\begin{equation}
H= ({\bf p}- q/c {\bf A} )^2/(2m) + V({\bf x}),
\end{equation}
 where $q$ is the electron charge, c the velocity of light,
 and ${\bf A}$ the vector potential due to an external magnetic field 
${\bf B}$.  
$V({\bf x})$ is a Gaussian distributed random function
\begin{equation}
< V({\bf x}) > = 0, < V({\bf x}) V({\bf x'}) > = \frac{\Delta}{ \tau} 
\frac{\delta ({\bf x} - {\bf x'})}{2 \pi S L},  
\end{equation}
 which models randomly distributed, uncorrelated impurities in the wire.
 
 The ASD 
 can be calulated for such a Hamiltonian 
as a function of the length of the wire 
 $ L$ and energy difference $\omega$.

 The unitary limit is considered, where the magnetic flux through
 the wire, $\phi$, exceeds $ \sqrt{ L/\xi } \phi_0$.

  The ASD can be written in terms of Grassmannian functional integrals.
 This allows to perform the impurity averaging as a Gaussian integral. 
 The resulting interacting theory of Grassman fields can be decoupled 
 by a transformation, introducing a functional integral over a 
$2 \times 2$-matrix Q. Next, the Grassmann fields can be integrated out. 
 The integral over Q can be simplified 
  for $\omega < 1/\tau$ 
 to an integral over gapless fluctuations
 around the saddle point which have the action of an O(3) nonlinear 
 sigma model
\begin{equation}
F[Q] = \frac{\pi}{2} \int \frac{d x}{S L} [ g Tr ( \frac{\nabla}{\pi/L} Q )^2
+ i \frac{\omega}{\Delta} Tr \Lambda Q ],
\end{equation}
 with the nonlinear constraint $Q^2 = 1$.   
Here, $ g = {   E_c/ \Delta } = \xi/L $
 which has the physical meaning of the
 dimensionless conductance $g= G/(e^2/h)$ of the wire,
 as long as the Einstein relation  
to
 the diffusion constant $D$  holds, $G = (\pi/4) e^2 \nu (S/L) D $.
 The derivation is given in \onlinecite{me}.
 A nonlinear sigma model for disordered electron systems had been 
 first derived in \cite{weg} for N replicas using a functional integral
over conventional numbers,  in
\cite{elk} for Grassmann variables, and then for superfields
\cite{ef}. 

 Choosing a representation of the matrix $Q$, the integrals can be performed
 by means of the transfer matrix method. Thus,
the problem  can be  reduced
 to the solution of the  equation\cite{me},
\begin{equation}\label{de} 
- L g \frac{d}{d z} P_z ( \lambda ) = \hat{H} [ \lambda ] P_z ( \lambda ), 
\end{equation}
  with the boundary condition $ P_L ( \lambda ) =1$. 
Here, $-1~<~\lambda~<~1$, $0~<~z~<~L$, and 
  the   Hamilton operator is,
\begin{equation}
 \hat{H} [\lambda] = - i \pi g \frac{ \omega}{\Delta} \lambda + \frac{\pi}{2 }
  \partial_{\lambda} ( 1 - \lambda^2 ) \partial_{\lambda}.
\end{equation}
 
 The ASD is then given by
\begin{equation}
C( \omega ) = \frac{1}{2} \int_{-1}^1 d \lambda P_0 ( \lambda ).
\end{equation} 
While we did not succeed to find an exact analytical solution of this 
 initial value problem, the function
\begin{eqnarray}
&& P_x ( \lambda ) = \exp [ i \frac{\omega}{ \Delta} g \lambda ( \exp  ( \pi 
 ( \frac{x}{L} - 1 )/g ) - 1)] \nonumber \\
 && \exp [ \frac{\pi}{2} (\frac{\omega }{\Delta } )^2 g^2 \int_{1/g}^{x/(L g)} d s
 ( \exp ( \pi ( s - 1/g ) ) -1 )^2 ] 
\end{eqnarray}  
 is a good approximation  when
 $ \omega^2  <  \Delta^2 g$, 
 for arbitrary $g$,
and becomes exact for $ g \rightarrow 0$
 when $ \omega^2 > \Delta^2/g $.

 Thus,  the ASD is obtained as
\begin{equation} \label{result}
 C( \omega ) = \frac{ \sin ( A_g \pi \omega/\Delta )}{ A_g \pi 
\omega/\Delta } \exp ( - B_g (\frac{ \omega}{\Delta})^2 ). 
\end{equation} 
 with 
\begin{equation}
A_g = \frac{g}{\pi} ( \exp ( - \frac{\pi}{g} ) - 1 ) ,
\end{equation}
 and
\begin{equation}
 B_g = \frac{g^2}{4} ( - \exp ( - \frac{2 \pi}{g} ) + 4 \exp ( - \frac{\pi}{g}
 ) - 3 + \frac{2 \pi}{ g} ). 
\end{equation}

 Fig. 2 
 shows a  plot of 
 the ASD as a function of the scaled frequency 
$x$ and the scaled length $t = L/\xi = 1/g   $.
A  clear damping of the amplitude of oscillations
accompanied by a shift of their phase is seen.

 This 
 shows that there is an  effect of localization on level correlations.
 At smaller $g$, the oscillations are damped more strongly,
  and  the envelope approaches a Gaussian decaying function.

 \begin{figure}[bhp]\label{fig2}
\epsfxsize=9cm
\centerline{\epsffile{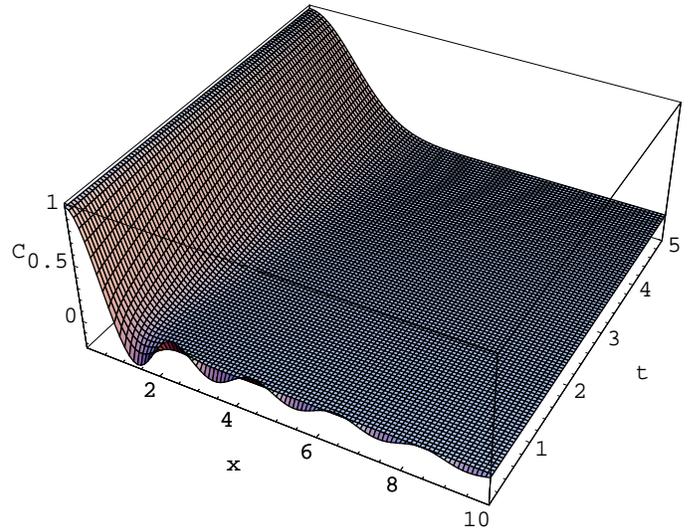}}
\caption{The ASD  
 as a function of scaled frequency $x = \omega/\Delta$,
 and  the  parameter 
$ t = 1/g = L/\xi$. } 
\end{figure}    

 To see this in more detail,
 let us  consider  approximations of Eq. (\ref{result}) 
 in  various regimes of interest.

 1.  In the metallic regime $ g > 1$ 
for $ x \ll g $ we  obtain  
:
\begin{equation}\label{met}
C(x ) = \frac{ \sin ( \pi x )}{\pi x} \exp ( - \frac{\pi^3}{6} \frac{1}{g} 
 x^2),
\end{equation} 
where $x = \omega/\Delta$,
 which for $ g \rightarrow \infty$ 
 reduces to the result  obtained with
 the unitary Wigner-Dyson ensemble of random
 matrices\cite{us}.

 2. In the  strongly localized regime $ g \ll 1$,
 one obtains:
\begin{equation} \label{poi}
C( \omega ) = \frac{ \sin ( g x )}{ g 
x } 
\exp ( - \frac{g}{2 } \pi  x^2 ).
\end{equation}
 Rescaling $ \tilde{x} = (g/\pi) \omega/\Delta= (1/\pi) \omega/\Delta_c$, we 
note the similarity to Eq. (\ref{met}). 

This result shows clearly that 
   the   correlations 
 between  energy levels belonging to states 
 which are spatially separated by more than the localization length 
 $\xi$ are  weak.
 As a result, the 
  ASD shows only correlations with the  period of twice  
 the  effective local energy level spacing
 $\pi~\Delta_c$  
of energy levels whose wave functions overlap spatially.
  
  As  $g \rightarrow 0$, the function is dominated by the 
 Gaussian factor.
 Thus, $C(\omega) =0$ exactly as $g\rightarrow 0$,
 and $\omega^2~>~\Delta^2/g$.

In Fig. 3 a plot of 
\begin{eqnarray}
&&F(t) = C( \Delta_c )= \frac{ \sin( \exp ( - \pi t ) - 1 )}{ \exp ( - \pi t)
-1 } 
\nonumber \\
&&   \exp ( - \frac{1}{4} ( - \exp ( - 2 \pi t ) + 4 \exp ( - \pi t ) - 3 + 2
\pi t ) ), 
\end{eqnarray}
is shown,
 where $ t= 1/g = L/\xi$.
  The ASD is decaying 
from $1$ to $0$ 
 as the  frequency  is held constant at $\omega = \Delta_c$ 
 and one varies the length of the wire $L$ or 
  the parameter $ t$ in Fig.~3, compare with Fig.~1.
 
\begin{figure}[bhp]\label{fig3}
\epsfxsize=9cm
\centerline{\epsffile{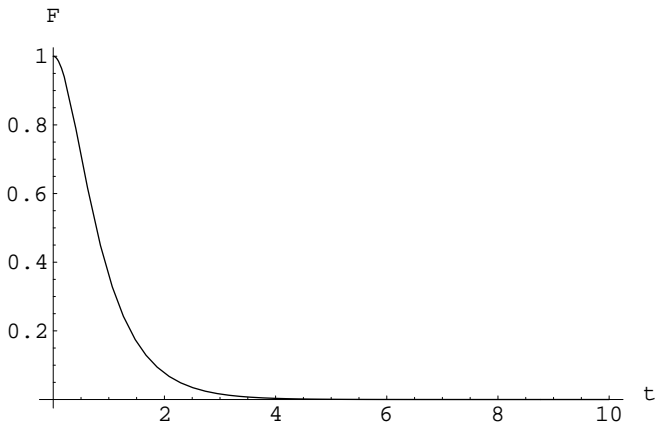}}
\caption{  $F(t)$,
 as a function of   the  scaled length of the wire 
$ t = 1/g = L/\xi$. } 
\end{figure}      
               
\section{Discussion} 
 
 In summary,   the ASD  is established as a  tool to 
 study the level statistics of disordered metals.
 An analytical
 expression is obtained  
 for the ASD of a quasi-1-dim. disordered mesoscopic wire.
 At frequencies below the mean level spacing $\Delta$  
 the ASD  approaches  1 like a Gaussian for any value of the 
conductance g,
 and there is no information on localization in this regime. 
This was pointed out by Efetov \cite{ef}
 when  studying the weakening of level repulsion
by localization. It was  stressed there, that the noncompact degrees
 of freedom are needed to describe localization that way.
 Here, it is shown that the information 
  is rather contained in the 
 large frequency correlations. The ASD 
 shows a crossover to a strong damping of the correlations 
 as the length of the wire exceeds the localization length $\xi$, 
 accompanied by the convergence of 
  the period of the strongly 
damped oscillations  to the constant $ 2 \pi \Delta_c$.
 Thus,  the wire can  be thought of as 
 effectively separated  into localization volumes, 
 as obtained earlier in Refs.
 \onlinecite{ke,alt}.

 One may argue that, 
 since the averaging
 over the impurity potential  was done before normalization,
 the resulting function might contain different information than
   the one obtained by  normalizing  for a given impurity potential
 before doing the averaging\cite{zile}. 
  The goal of  this article 
 is however  to show that
  level statistics can be characterized with the simplest 
 tool, $C(\omega)$.

 Now, it might become possible  to address 
 problems analytically, which could not be solved with the  methods 
 known so far, due to their complexity. 
 While the ASD cannot contain any information on the 
 eigenfunctions of the system,
 we have seen that 
 it 
contains enough information to characterize the energy level statistics.

 The function $F(t)$  may serve as a
  parameter characterizing localization~: it is $1$ in the metallic regime
 and $0$, when all states at all energies are localized.   
  It decays to approximately $1/e$ when 
 the length of the wire coincides with the localization length, $ L= \xi$.
      
 In addition, recently it has been shown that the ASD can
 contain information not only on a metal-insulator-crossover,
 but also on a -transition, 
 as demonstrated with 
 the Anderson model on a Bethe lattice\cite{me}.

 The author would like to thank Uzy Smilansky for drawing his
 attention to the ASD as a tool to study level statistics,
 and Thomas Dittrich, 
Konstantin Efetov, Dietrich Klakow, D. E. Khmel'nitskii, Igor Lerner, 
 Daniel Miller, Vladimir Prigodin
 and Klaus Ziegler
   for 
 usefull discussions, and Simon Villain- Guillot for critical reading 
 of the manuscript. 
  
  This work was possible thanks to a scholarship by Minerva
 and   support by the Max Planck
 Institute of Physics of Complex Systems in Dresden.

\end{document}